\documentclass[reprint,aps,superscriptaddress,notitlepage]{revtex4-1}
\usepackage{amssymb,amsmath}
\usepackage{graphicx}
\usepackage{dcolumn}
\usepackage{multirow}
\usepackage{color}
\usepackage{bm}
\usepackage[english]{babel}
\usepackage[caption=false]{subfig}

\newcommand{\beginsupplement}{%
        \setcounter{table}{0}
        \renewcommand{\thetable}{S\arabic{table}}%
        \setcounter{figure}{0}
        \renewcommand{\thefigure}{S\arabic{figure}}%
        \setcounter{equation}{0}
        \renewcommand{\theequation}{S\arabic{equation}}
        \setcounter{section}{0}
        \renewcommand{\thesection}{S\Roman{section}}
        }
\begin{document}

\renewcommand{\figurename}{FIG.}

\title{Measurements of capacitive coupling within a quadruple quantum dot array}
\author{Samuel F. Neyens}
\author{E. R. MacQuarrie}
\author{J. P. Dodson}
\author{J. Corrigan}
\author{Nathan Holman}
\author{Brandur Thorgrimsson}
\author{M. Palma}
\author{Thomas McJunkin}
\affiliation{University of Wisconsin-Madison, Madison, WI 53706, USA}
\author{L. F. Edge}
\affiliation{HRL Laboratories, LLC, 3011 Malibu Canyon Road, Malibu, CA 90265, USA}
\author{Mark Friesen}
\affiliation{University of Wisconsin-Madison, Madison, WI 53706, USA}
\author{S. N. Coppersmith}
\affiliation{University of Wisconsin-Madison, Madison, WI 53706, USA}
\affiliation{University of New South Wales, Sydney, Australia}
\author{M. A. Eriksson}
\affiliation{University of Wisconsin-Madison, Madison, WI 53706, USA}

\begin{abstract}
We present measurements of the capacitive coupling energy and the inter-dot capacitances in a linear quadruple quantum dot array in undoped Si/SiGe. With the device tuned to a regime of strong ($>$1 GHz) intra-double dot tunnel coupling, as is typical for double dot qubits, we measure a capacitive coupling energy of $20.9 \pm 0.3$ GHz. In this regime, we demonstrate a fitting procedure to extract all the parameters in the 4D Hamiltonian for two capacitively coupled charge qubits from a 2D slice through the quadruple dot charge stability diagram. We also investigate the tunability of the capacitive coupling energy, using inter-dot barrier gate voltages to tune the inter- and intra-double dot capacitances, and change the capacitive coupling energy of the double dots over a range of 15-32 GHz. We provide a model for the capacitive coupling energy based on the electrostatics of a network of charge nodes joined by capacitors, which shows how the coupling energy should depend on inter-double dot and intra-double dot capacitances in the network, and find that the expected trends agree well with the measurements of coupling energy. 
\end{abstract}

\maketitle

\section{Introduction}

Electron spins in semiconductor quantum dots are a promising platform for quantum computation \cite{Loss:1998p120,Russ:2017p393001,Zhang:2018p32}. Quantum dots formed in Si/SiGe heterostructures have many advantages, including high electron mobility, low natural abundance of spinful isotopes in Si, and compatibility with industrial Si-based fabrication techniques \cite{Zwanenburg:2013p961}. Such devices were initially realized in doped heterostructures \cite{Klein:2004p4047}, but the transition to undoped, fully gated structures \cite{Lu:2011p043101,Borselli:2011p063109,Wang:2013p046801,Wu:2014p11938} has led to improved charge stability. Undoped Si/SiGe heterostructures have now hosted many qubit architectures, with recent demonstrations of single dot qubits such as the Loss-DiVincenzo qubit \cite{Kawakami:2016p11738,Yoneda:2018p102,Zajac:2018p439,Watson:2018p633,Mi:2018p599,Xue:2019p021011}; double dot qubits such as the singlet-triplet qubit \cite{Maune:2012p344,Wu:2014p11938,Reed:2016p110402}, quantum dot hybrid qubit \cite{Kim:2014p70,Thorgrimsson:2017p32}, and valley qubit \cite{Schoenfield:2017p64,Mi:2018p161404}; and triple dot qubits such as the exchange-only qubit \cite{Eng:2015p4}.

Two-qubit gates in semiconductor quantum dots have been demonstrated through use of the exchange coupling in Loss-DiVincenzo qubits \cite{Nowack:2011p1269,Brunner:2011p146801,Veldhorst:2015p410,Zajac:2018p439,Watson:2018p633,Xue:2019p021011,Hendrickx:2019preprint} and through use of the capacitive coupling in singlet triplet qubits \cite{Shulman:2012p202,Nichol:2017p1} and charge qubits \cite{Li:2015p7681}. For double dot qubits, the capacitive interaction arises when the individual qubit states, $|0\rangle$ and $|1\rangle$, have different admixtures of the eigenstates of electron position, $|L\rangle$ and $|R\rangle$. This difference can be described as an effective dipole moment for each qubit, leading to a dipole-dipole interaction between the qubits. The maximum such interaction energy between two double dot qubits is equal to the shift in detuning experienced by one double dot qubit due to the complete transfer of an electron between dots in the neighboring qubit. This interaction energy can be obtained by measuring the shift in the polarization line of one double dot due to a change in polarization of the other double dot \cite{VanWeperen:2011p030506,Yu:2016p324003,Ward:2016p16032,Zajac:2016p054013}. The resulting energy shift is the coupling term, $g$, in the Hamiltonian for two double dot qubits that interact capacitively. We refer to this energy from here on as the capacitive coupling.

In this work, we report measurements of the capacitive coupling in a quadruple quantum dot device in undoped Si/SiGe. We tune the device to a regime of strong intra-double dot tunnel coupling ($t > 1$ GHz in both double dots), to match the conditions of typical double dot qubit experiments, and measure the capacitive coupling to be $20.9 \pm 0.3$ GHz. In this regime, we demonstrate a fitting procedure with which we obtain, from a 2D slice through the quadruple dot charge stability diagram, all the parameters in the 4D Hamiltonian for capacitively coupled charge qubits. We investigate the tunability of the capacitive coupling \emph{in situ}, using barrier gate voltages to change the inter-double dot (inter-DD) and intra-double dot (intra-DD) capacitances in the quadruple dot array, and find the capacitive coupling changes over a range of 15-32 GHz, in a way that trends positively with inter-DD capacitance and negatively with intra-DD capacitances. We interpret the range of inter-dot capacitances observed here in terms of changes in inter-dot spacing, and estimate that the changes we make to the barrier gate voltages shift the positions of the quantum dots by tens of nm. We provide a simple model based on the electrostatics of a system of charge nodes joined by capacitors to illustrate how the capacitive coupling should depend on inter-DD and intra-DD capacitances, and we find that the expected trends from the model agree well with the trends in the measured data.

\begin{figure*}[t]
\includegraphics[width=1.0\textwidth]{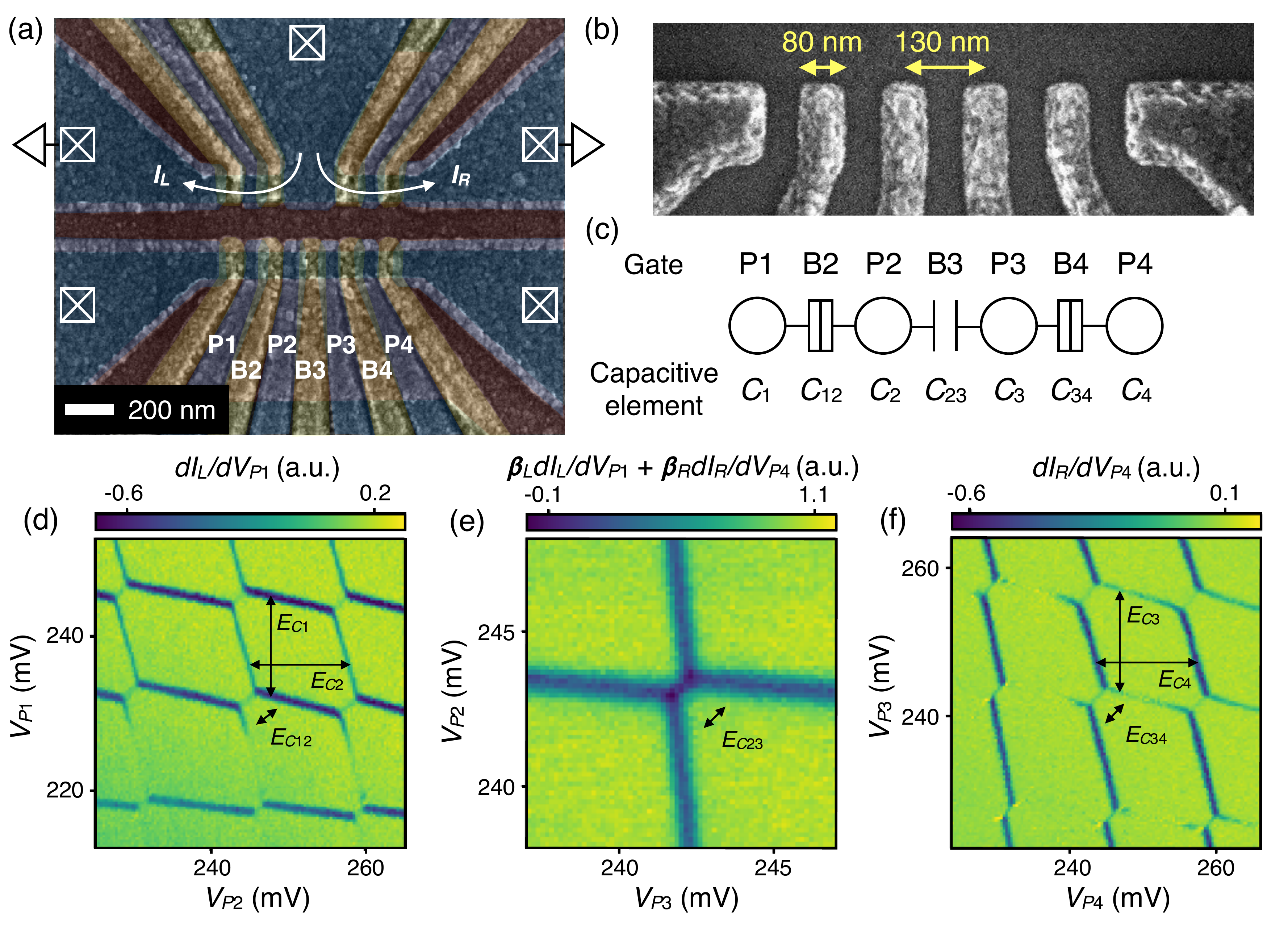}
\caption{
(a) False-colored SEM micrograph of a device lithographically identical to the device measured here. Plunger (P) and barrier (B) gates used in the tuning of the 4-dot array are labeled. The charge sensor currents $I_{L}$ and $I_{R}$ are also labeled. (b) Higher magnification image of the accumulation layer pattern showing the dimensions of the plunger gate array. (c) Schematic of the model of the four-dot array as a network of charge nodes and capacitors. $C_{i}$ is the sum of capacitances to dot $i$, and $C_{ij}$ is the capacitance between dots $i$ and $j$. Capacitances to reservoirs and gates are not shown. (d)-(f) Charge stability diagrams for each nearest-neighbor pair of dots in the array. The indicated energy scales correspond to the dot charging energies $E_{Ci}$ and the electrostatic coupling energies $E_{Cij}$. $E_{Ci}$ range from 2.4 to 4.4 meV, and $E_{Cij}$ range from 120 to 680 $\mu$eV, depending on the tuning. The data shown in (e) are a weighted sum of transconductance signals from the two charge sensor amplifiers: $\beta_{L} dI_{L} / dV_{P1} + \beta_{R} dI_{R} / dV_{P4}$, where $\beta_{L(R)}^{-1}$ is the range of the signal from the left (right) amplifier. Here, $\beta_{L} = 1.9 \times 10^{4}$ and $\beta_{R} = 78$. The difference in signal range is due to the difference in amplification schemes for the two charge sensors, as described in the main text.
}
\label{fig:fab_meas}
\end{figure*}

\section{Results and Discussion}

\subsection{Fabrication and measurement}

The device we study is composed of six quantum dots, four arrayed linearly in the main channel and two in separate channels used to sense the electron occupation of the array. A false-colored SEM micrograph of a lithographically identical device is shown in Fig.~\ref{fig:fab_meas}(a). The device is an accumulation-mode overlapping gate device with three layers, one each for screening, accumulation, and tunnel barrier control (see Suppl. Mat. for further details \cite{SM}). Quantum dot chemical potentials and inter-dot barrier potentials are primarily controlled by the plunger (P) and barrier (B) gates, respectively. The plunger gates are 80 nm wide with 130 nm pitch, as shown in a higher magnification image of the accumulation layer pattern in Fig.~\ref{fig:fab_meas}(b).

Measurements are performed in a dilution refrigerator with a base temperature below $20$ mK. The device is tuned to form a quantum dot under each plunger gate in the main channel, resulting in a linear array of four quantum dots. Two quantum dots are also formed in the auxiliary channels as charge sensors,  with the left (right) charge sensor mostly sensitive to double dot 1-2 (3-4). The left charge sensor is connected to a cryogenic amplifier similar to that in Ref. \cite{Tracy:2016p063101}; the right charge sensor current is amplified only at room temperature. Measurements of the charge occupation of the four-dot array are performed by modulating a plunger gate above each double dot and measuring the charge sensor currents with lock-in amplifiers at those modulation frequencies. The quantum dots are set to desired electron occupations by finding the last electron transitions in the dots and then counting up on the charge stability diagram. The tunnel couplings between dots are controlled with barrier gate voltages. $V_{B2}$ and $V_{B4}$ are generally tuned to be much more positive than $V_{B3}$, so that the tunnel couplings within each double dot are large while the tunnel coupling between the double dots is negligible. Thus, the significant coupling between the double dots is capacitive.

\begin{figure}[t]
\includegraphics[width=0.48\textwidth]{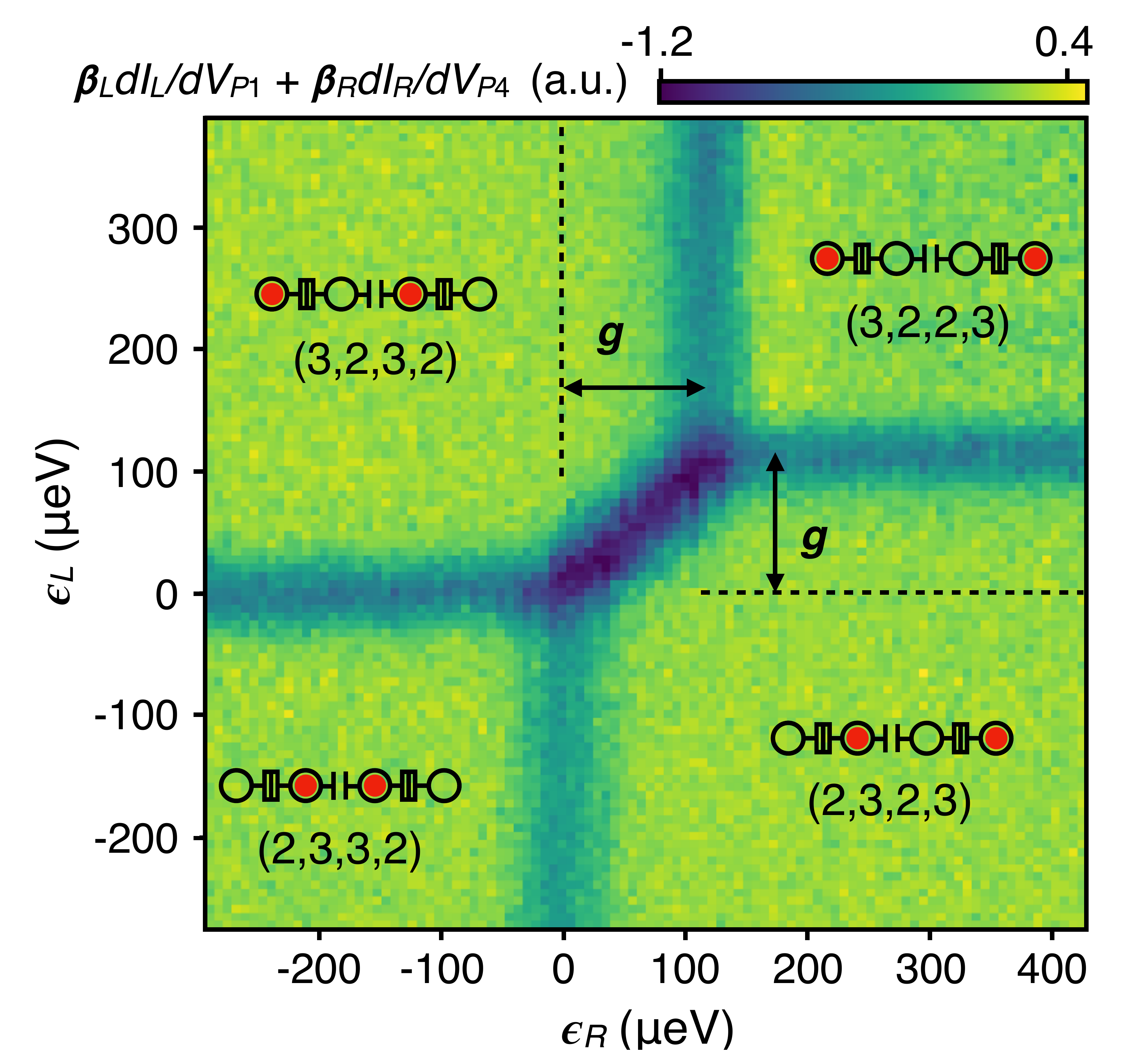}
\caption{
Measurement of the capacitive coupling energy between two double dots from the shift of the polarization line. The arrows indicate the magnitude of the polarization line shift, and the dashed lines indicate where the lines would be for $g=0$. For this dataset, $g= 28.4 \pm 0.4$ GHz.  $V_{P1}$ ($V_{P4}$) is used to sweep $\epsilon_{L}$ ($\epsilon_{R}$). The data shown is a weighted sum of transconductance signals from the two charge sensor amplifiers: $\beta_{L} dI_{L} / dV_{P1} + \beta_{R} dI_{R} / dV_{P4}$, with $\beta_{L} = 1.6 \times 10^{4}$ and $\beta_{R} = 68$.
}
\label{fig:g_meas}
\end{figure}

Fig.~\ref{fig:g_meas} shows a measurement of the capacitive coupling between the double dots. This measurement is done by measuring a 2D slice through the 4D quadruple dot charge stability diagram. By simultaneously sweeping the detuning of both double dots, we observe the shift in the polarization line of each double dot due to the change in polarization of the other double dot. The magnitude of the shift is extracted by fitting line cuts of each polarization line, finding the center point in each line cut, and fitting the curve describing the shift of the center points in detuning space. The functional form used to fit this curve is a hyperbolic tangent (tanh), based on the expected form for the polarization of a double dot as a function of its detuning \cite{Dicarlo:2004p1440}. The amplitude of this tanh function in units of detuning energy, indicated by the arrows in Fig.~\ref{fig:g_meas}, is equal to the capacitive coupling, $g$, which for the measurement in Fig.~\ref{fig:g_meas} is found to be $28.4 \pm 0.4$ GHz.

At each tuning of the barrier gate voltages, the detuning lever arms for both double dots are measured by sweeping the temperature and measuring the broadening of the polarization lines \cite{Simmons:2009p3234}, which also enables an extraction of the electron temperature $T_{e}=155$ mK. We also measure the individual capacitive elements of the quadruple dot system, shown schematically in Fig.~\ref{fig:fab_meas}(c), including $C_{i}$, the total capacitance to dot $i$, and $C_{ij}$, the capacitance between dots $i$ and $j$. These capacitances are obtained from the corresponding self-charging energies $E_{Ci}$ and electrostatic coupling energies $E_{Cij}$, which can be read from the charge stability diagrams by the dimensions labeled in Fig.~\ref{fig:fab_meas}(d)-(f) \cite{SM}.

The double dots are both tuned to be near the (3,2)-(2,3) polarization line, which is the charge qubit regime with the first valley shell filled in all dots. Using this electron configuration enables the detection of transitions of the inner two dots (2 and 3), which is necessary to measure the electrostatic energies indicated in Fig.~\ref{fig:fab_meas}(d)-(f). Since these dots are not directly coupled to reservoirs, their transitions require cotunneling through the outer dots (1 and 4), the rate of which becomes suppressed when the outer dots are empty and their chemical potentials lie well above the chemical potentials of the inner dots \cite{Nazarov:2009}.

\subsection{Capacitive coupling at strong inter-dot tunnel couplings}

The capacitive coupling measurement shown in Fig.~\ref{fig:g_meas} is taken with low intra-DD tunnel couplings. From lack of tunnel broadening of the polarization lines, we determine $t_{12},t_{34} < k_{B}T_{e}$ for that measurement, where $k_{B}T_{e} \sim 3$ GHz. Suitable values of the tunnel couplings and the capacitive coupling are important for enabling high-fidelity single- and two-qubit gates. To enable good single-qubit control, the intra-DD tunnel couplings should typically be on the order of $\sim$1-10 GHz between dots in charge \cite{Petersson:2010p246804,Shi:2013p075416,Kim:2015p243}, singlet-triplet \cite{Maune:2012p344,Wu:2014p11938}, quantum dot hybrid \cite{Kim:2014p70,Cao:2016p086801,Thorgrimsson:2017p32}, and valley \cite{Schoenfield:2017p64,Mi:2018p161404} qubits. Furthermore, to couple the qubits purely capacitively, the tunnel rate between dots 2 and 3 should be low so that the exchange coupling between dots 2 and 3 is negligible and the probability of state leakage across B3 during control and readout of the qubits is low.

Taking these considerations into account, we look at an example configuration with strong intra-DD tunnel coupling and weak inter-DD tunnel coupling and measure the capacitive coupling of the system. We set $V_{B3}$ to achieve a low inter-DD tunneling rate $t_{23}\lesssim 1$ kHz, measured by varying the lock-in frequency and tracking the visibility of the polarization line between dots 2 and 3. We raise $V_{B2}$ and $V_{B4}$ until $t_{12},t_{34} > k_{B}T_{e}$, determined by observing tunnel broadening of the intra-DD polarization lines. In this regime, we measure $g = 20.9 \pm 0.3$ GHz. This is reduced compared to the measurement with weaker intra-DD tunnel coupling shown in Fig.~\ref{fig:g_meas}, where $g = 28.4 \pm 0.4$ GHz, but is still expected to be strong enough to perform high fidelity two-qubit gates for quantum dot hybrid qubits \cite{Frees:2018preprint}.

\begin{figure}[t]
\includegraphics[width=0.48\textwidth]{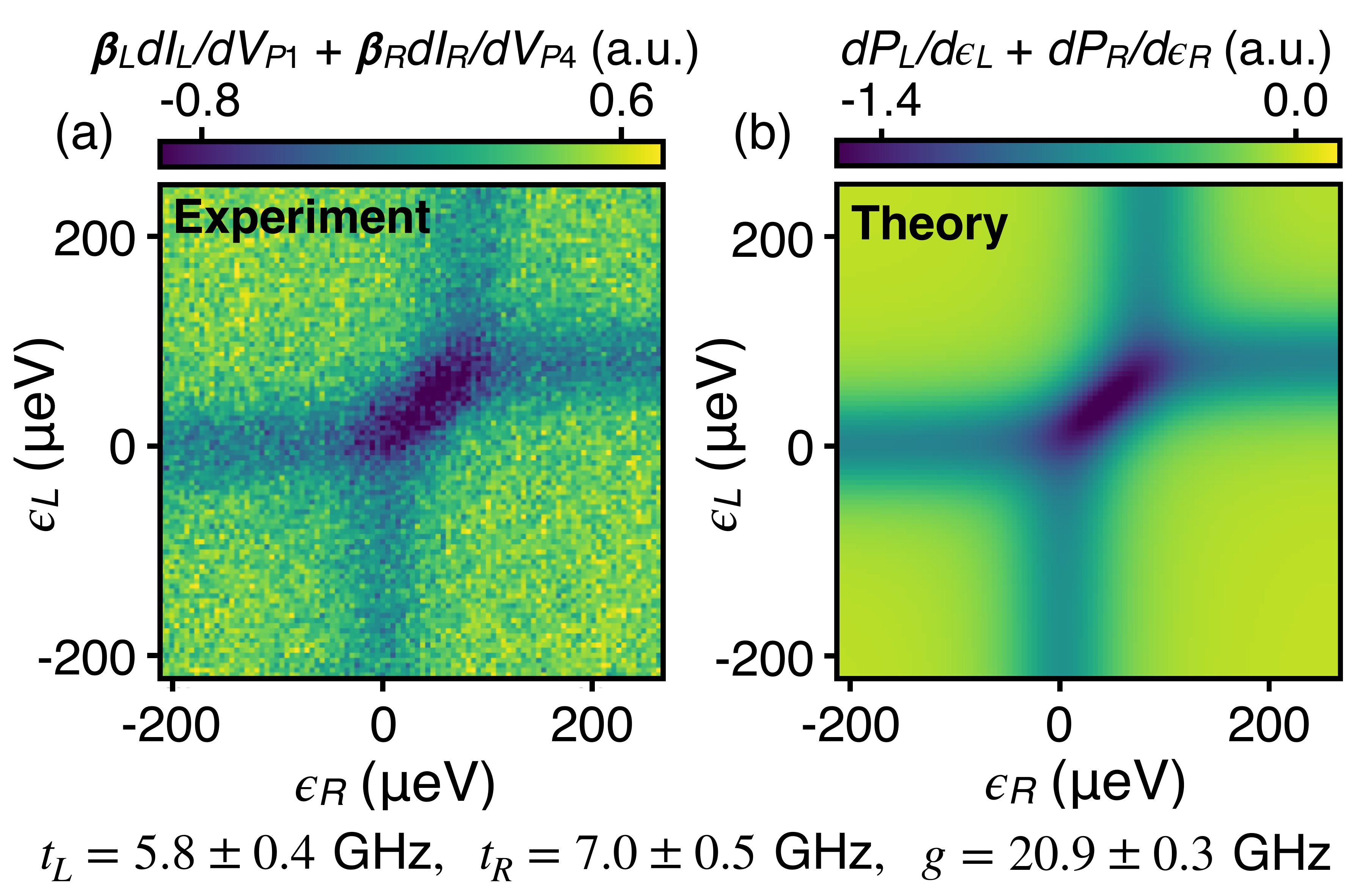}
\caption{
(a) A 2D slice through the quadruple dot stability diagram, taken using $V_{P1}$ ($V_{P4}$) to sweep $\epsilon_{L}$ ($\epsilon_{R}$), with $t_{23} \lesssim 1$ kHz and $t_{12},t_{34} > k_{B}T_{e}$. The data shown is a weighted sum of transconductance signals from the two charge sensor amplifiers, with $\beta_{L} = 4.0 \times 10^{4}$ and $\beta_{R} = 1.4 \times 10^{2}$. (b) A theoretical fit to the data based on the 4D Hamiltonian for two coupled charge qubits. Hamiltonian parameters extracted from the fit are listed at bottom.
}
\label{fig:g_meas_theory}
\end{figure}

Fig.~\ref{fig:g_meas_theory}(a) shows a 2D slice of the quadruple dot stability diagram at this configuration. As explained above, the shift of the polarization lines in energy corresponds to the magnitude of the capacitive coupling. Additionally, at this tuning where intra-DD tunnel couplings are high, each shifted polarization line acquires an increased curvature due to the tunnel broadening of the opposite polarization line. By adopting a more sophisticated model that incorporates the effects of tunnel coupling, electron temperature, and capacitive coupling, we can fit this curvature to extract more information about the Hamiltonian of the quadruple dot system. Using this analysis, the 2D dataset shown in Fig.~\ref{fig:g_meas_theory}(a) yields all the parameters in the Hamiltonian for two coupled charge qubits in the absence of noise, given a known detuning lever arm. The procedure is as follows. We write the 4D Hamiltonian,
\begin{multline}
H = \frac{\epsilon_{L}}{2} \sigma_{z} \otimes I + t_{L} \sigma_{x} \otimes I + \frac{\epsilon_{R}}{2} I \otimes \sigma_{z} + t_{R} I \otimes \sigma_{x} \\
+ \frac{g}{4} (I - \sigma_{z}) \otimes (I - \sigma_{z}),
\end{multline}
where $\epsilon_{L(R)}$ is the detuning in the left (right) double dot, $t_{L(R)} = t_{12(34)}$, $g$ is the capacitive coupling, $I$ is the identity operator, and $\sigma_{i}$ are the usual Pauli operators. From $H$, we obtain the eigenstates $| \psi_{i} \rangle$ as functions of $\epsilon_{L}$, $\epsilon_{R}$, $t_{L}$, $t_{R}$, and $g$. Then, extending the method in Ref.~\cite{Dicarlo:2004p1440} to a two-qubit system, we calculate the expectation value of the charge polarization of each double dot, averaged over a Maxwell-Boltzmann distribution:
\begin{multline}
P_{L(R)}(\epsilon_{L}, \epsilon_{R}; t_{L}, t_{R}, g) = \\
 \frac{1}{Z} \sum_{i=1}^{4} \langle \psi_{i} | \sigma_{z}^{L(R)} | \psi_{i} \rangle e^{-E_{i} / k_{B} T_{e}},
\end{multline}
where $\sigma_{z}^{L} = \sigma_{z} \otimes I$, $\sigma_{z}^{R} = I \otimes \sigma_{z}$, and Z is the partition function. This expression yields two functions, one for the charge polarization of each double dot as a function of $\epsilon_{L}$ and $\epsilon_{R}$ and parametrized by $t_{L}$, $t_{R}$, and $g$. Fitting each of these functions to the shifted polarization line data in Fig.~\ref{fig:g_meas_theory}(a) yields the theoretical stability diagram shown in Fig.~\ref{fig:g_meas_theory}(b) \cite{SM}. From the fit we extract the parameters $t_{L} = 5.8 \pm 0.4$ GHz, $t_{R} = 7.0 \pm 0.5$ GHz, and $g = 20.9 \pm 0.3$ GHz, allowing us to write out the complete 4D Hamiltonian for coupled charge qubits at every point in Fig.~\ref{fig:g_meas_theory}(a).

\subsection{Controlling the capacitive coupling with barrier gate voltages}

Comparison of the measurements in Fig.~\ref{fig:g_meas} and Fig.~\ref{fig:g_meas_theory}(a) shows how a change of barrier gate voltages that increases the intra-DD tunnel couplings results in a significant decrease in the capacitive coupling ($\sim25\%$). We further investigate the tunability of the capacitive coupling in response to the barrier gate voltages $V_{B2}$, $V_{B3}$, and $V_{B4}$, by measuring the coupling energy as well as all the parameters of the capacitance network shown in Fig.~\ref{fig:fab_meas}(c) as a function of these voltages. 

\begin{figure}
\includegraphics[width=0.48\textwidth]{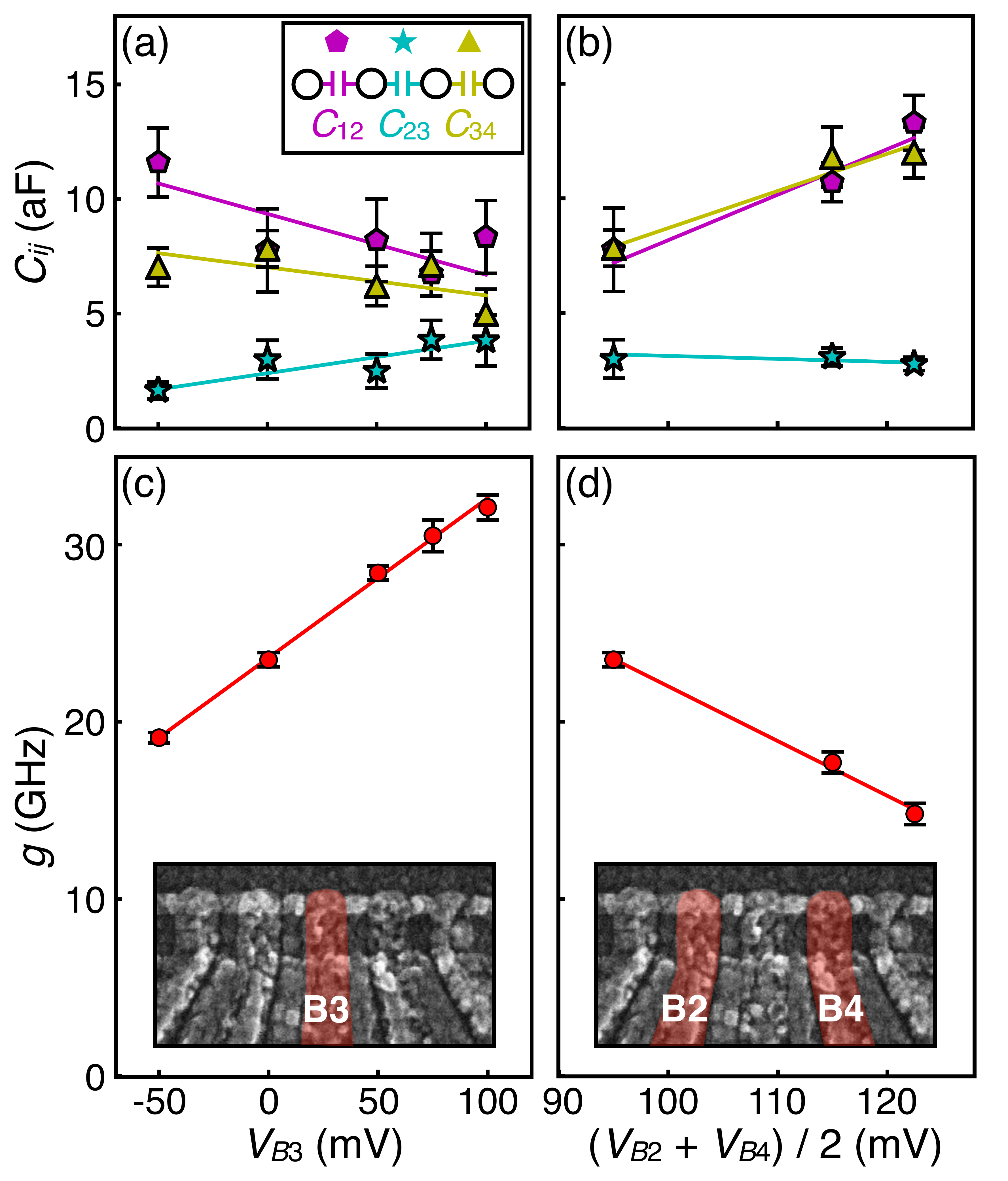}
\caption{
(a) Inter-dot capacitances as a function of the middle barrier gate voltage, $V_{B3}$, with $(V_{B2},V_{B4}) = (120,70)$ mV. (b) Inter-dot capacitances as a function of the average of the two outer barrier gate voltages, $V_{B2}$ and $V_{B4}$, with $V_{B3} =$~0 mV. The gate voltage values for all points are, from left to right: $(V_{B2},V_{B4}) = (120,70)$, (140,90), and (155,90) mV. (c) Capacitive coupling as a function of $V_{B3}$. Inset: false-colored SEM micrograph highlighting the gate whose voltage is varied. (d) Capacitive coupling as a function of $V_{B2}$ and $V_{B4}$. Inset: false-colored SEM micrograph highlighting the gates whose voltages are varied. For all plots, linear fits to the data are shown as a guide for the eye.
}
\label{fig:g_C_tuning}
\end{figure}

Fig.~\ref{fig:g_C_tuning} shows the results of the measurements, where we observe a range of capacitive couplings from 15-32 GHz in response to changes in inter-dot barrier gate voltages. Each of these barrier gate voltages tunes the capacitance between the dots straddling that barrier. In this way, we investigate the relationships among capacitive coupling, inter-DD capacitance, and intra-DD capacitances. In Fig.~\ref{fig:g_C_tuning}(a) and (c), the middle barrier voltage, $V_{B3}$, is varied with $V_{B2}$ and $V_{B4}$ held fixed. Panel (c) shows how increasing $V_{B3}$ increases the capacitive coupling, $g$. Panel (a) shows the effect of $V_{B3}$ on all the inter-dot capacitances in the system. Increasing $V_{B3}$ increases the inter-DD capacitance ($C_{23}$) and also decreases both intra-DD capacitances ($C_{12}$ and $C_{34}$). These changes in capacitance are a result of the dots' position shifting in the array. Making $V_{B3}$ more positive decreases the potential between dots 2 and 3, resulting in these dots shifting closer together (increasing $C_{23}$) and farther from their outer neighbors, 1 and 4 (decreasing $C_{12}$ and $C_{34}$). The effects of these changes in gate voltage on the coupling $g$ follow the intuition for a dipole-dipole interaction, where a decrease in the spacing between the dipoles causes an increase in the interaction energy.

In Fig.~\ref{fig:g_C_tuning}(b) and (d), $V_{B3}$ is held fixed and $V_{B2}$ and $V_{B4}$ are varied. Panel (d) shows how increasing $V_{B2}$ and $V_{B4}$ decreases $g$. Panel (b) shows the effect of these barrier voltages on the inter-dot capacitances. Here, the change in inter-DD capacitance is small, while the intra-DD capacitances change significantly in response to $V_{B2}$ and $V_{B4}$. Making these voltages more positive decreases the potential in the middle of each double dot, shrinking the spacing between dots in each pair (increasing $C_{12}$ and $C_{34}$). The effects of these changes in gate voltage on the coupling $g$ again follow the intuition for a dipole-dipole interaction, where here a decrease in intra-DD spacing corresponds to a decrease in the size of the dipoles, which decreases the interaction energy.

To estimate the shift in quantum dot positions associated with the changes in capacitance observed in Fig.~\ref{fig:g_C_tuning}(a) and (b), we model a pair of neighboring dots as two conducting discs beneath a conducting plane, which incorporates the screening effects from the overlapping gate metal \cite{SM}. We assume a dot diameter equal to the plunger gate width of 80 nm. The capacitance in this model follows an approximate $1/d^{3}$ dependence, where $d$ is the center-to-center distance between the dots. Varying $d$ from 85 to 175 nm, to cover the range over which this spacing could vary in an array of 80 nm dots with 130 nm gate pitch, we calculate inter-dot capacitances ranging from 1-10 aF, in good agreement with the measured capacitances in Fig.~\ref{fig:g_C_tuning}(a) and (b), which range from 2-13 aF. These numbers also suggest that the variations in inter-dot capacitance observed in Fig.~\ref{fig:g_C_tuning} are the result of significant shifts in dot position, on the order of tens of nm, with a dot pitch on the low $C_{ij}$ end ($\sim1$ aF) of $\sim170$ nm and a dot pitch on the high $C_{ij}$ end ($\sim10$ aF) of $\sim90$ nm. 

To further understand the contributions that the inter- and intra-DD capacitances make to the capacitance coupling, we model the quadruple dot system as a network of four charge nodes joined by capacitors. The capacitive coupling creates a detuning shift in one double dot due to the change in polarization of the other double dot. We extend the analysis from Ref. \cite{vanderWiel:2003p1} from two to four quantum dots and calculate this detuning shift to obtain an analytical expression for capacitive coupling as a function of the capacitive elements shown in Fig.~\ref{fig:fab_meas}(c) (details in Suppl. Mat. \cite{SM}):
\begin{equation}
g = \frac{e^{2}}{|\mathbf{C}|} C_{23}(C_{1}-C_{12})(C_{4}-C_{34})
\end{equation}
where $e$ is the electron charge and $|\mathbf{C}|$ is the determinant of the capacitance matrix. We can simplify the expression by approximating all $C_{i}=C$. Then, assuming inter-dot capacitances are small compared to total dot capacitances, we can series expand in the ratios $c_{ij}=C_{ij}/C$, finding, to second order:
\begin{equation}
g / E_{C} = c_{23} - c_{23}c_{12} - c_{23}c_{34},
\label{eq:gfunction}
\end{equation}
where $E_{C}=e^{2}/C$ is the single-dot charging energy. The approximate expression in Eq.~\ref{eq:gfunction} provides intuition for the relative contributions that the inter-DD and intra-DD capacitances make to the capacitive coupling and how the capacitive coupling should trend with each. The leading contribution of the inter-DD capacitance ($c_{23}$) is first order, while the leading contributions of the intra-DD capacitances ($c_{12}$ and $c_{34}$) are second order. The capacitive coupling depends positively on inter-DD capacitance but negatively on intra-DD capacitances. This agrees with the correlations we observe between inter-dot capacitances and capacitive coupling, as shown in Fig.~\ref{fig:g_C_tuning}, where changes in inter-DD (intra-DD) capacitance correlate positively (negatively) with changes in capacitive coupling.

\subsection{Discussion}

The results in Fig.~\ref{fig:g_C_tuning} demonstrate a large degree of control over the capacitive coupling using the inter-dot barrier voltages to change the inter-dot capacitances in the array. When this device is tuned to a realistic regime for performing two-qubit experiments, where intra-DD tunneling coupling is high ($t_{12},t_{34} > 1$ GHz) and the inter-DD tunnel coupling is very low ($t_{23} \lesssim 1$ kHz), we find a capacitive coupling of $\sim 20$ GHz, which corresponds to a fast 2-qubit entangling time of $\sim 20$ ps when both qubits have equal admixtures of $|L\rangle$ and $|R\rangle$ states. Based on the trends observed in Fig.~\ref{fig:g_C_tuning}(c), if an even higher capacitive coupling rate were desired, we expect the coupling could be increased further in this device by raising $V_{B3}$ to increase $C_{23}$ while raising $V_{B2}$ and $V_{B4}$ to maintain strong intra-DD tunnel couplings. This would raise the inter-DD leakage rate across B3, but for many semiconductor qubits, this rate could be brought into the MHz range without surpassing the operation rate of the qubits themselves. The ability to raise $C_{23}$ while keeping $t_{23}$ low could also be enhanced further by making the barrier potential between dots 2 and 3 higher and narrower, which could be achieved by a straightforward lithographic change of decreasing the gap between plunger gates P2 and P3. We note also that a proportional decrease of all distances in the dot array leads to increased values of $g$, which is consistent with the larger values of $g$ ($\sim50$ GHz) that have been measured in a dot array with a smaller (100 nm) pitch \cite{Zajac:2016p054013}.

\section{Conclusion}

We measure the capacitive coupling and all inter-DD and intra-DD capacitances in a linear array of four quantum dots in the few-electron regime at a range of tunings. We tune to a regime of strong intra-DD tunnel coupling and measure the capacitive coupling to be $g = 20.9 \pm 0.3$ GHz, which is strong enough to be able to implement high-fidelity two-qubit operations. We demonstrate a fitting procedure to extract all the parameters of the 4D Hamiltonian for two capacitively coupled charge qubits from a 2D slice through the quadruple dot stability diagram. We tune the capacitive elements in the quadruple dot array with inter-dot barrier gate voltages and see the capacitive coupling change over a range of 15-32 GHz. We provide a simple model based on a system of charge nodes joined by capacitors to illustrate how capacitive coupling should depend on the inter-DD and intra-DD capacitances of the system and find the model agrees well with the trends in the measured data. 

\section*{Acknowledgements}

Research was sponsored in part by the Army Research Office (ARO) under Grant Number W911NF-17-1-0274 and by the Vannevar Bush Faculty Fellowship program under ONR grant number N00014-15-1-0029. We acknowledge the use of facilities supported by NSF through the UW-Madison MRSEC (DMR-1720415). The views and conclusions contained in this document are those of the authors and should not be interpreted as representing the official policies, either expressed or implied, of the Army Research Office (ARO), or the U.S. Government. The U.S. Government is authorized to reproduce and distribute reprints for Government purposes notwithstanding any copyright notation herein.

\beginsupplement

\section*{Supplementary Information}

\section{Device fabrication and packaging}
The device is fabricated using undoped Si/SiGe heterostructures grown by chemical vapor deposition. A 3~$\mu$m thick linearly graded Si$_{0.7}$Ge$_{0.3}$ relaxed buffer substrate is grown on top of a Si wafer. The buffer is chemically and mechanically polished before the growth of a 170-375 nm thick Si$_{0.7}$Ge$_{0.3}$ layer, a 9 nm thick Si quantum well, a 30 nm thick Si$_{0.7}$Ge$_{0.3}$ spacer, and a 1 nm thick Si cap \cite{Deelman:2016p224}. The layers are measured using x-ray diffraction and x-ray reflectivity to confirm the composition and thickness respectively \cite{Richardson:2017p02B113}.

A mesa for each quantum dot device is etched using a CHF$_{3}$ plasma. Ohmic contacts are implanted with $^{31}$P$^{+}$ dopants and activated with an anneal. A dielectric stack of Al$_{2}$O$_{3}$ is deposited with atomic layer deposition in two steps to form a terraced structure with a 5 nm gate oxide in the region of quantum dot formation and a 20 nm field oxide over the bulk of the ohmic reservoirs. After deposition, the oxide is annealed in forming gas at 450$^{\circ}$C for 15 min. Bond pads, interconnects, and ohmic metal are made with Pd deposited in an e-beam evaporator. The bond pads for each device are made with metallic shorting lines to protect against electrostatic discharge (ESD) during subsequent e-beam lithography and wire bonding. The quantum dot gates are fabricated with e-beam lithography, using PMMA 495 A4 resist and depositing Al in an e-beam evaporator. After liftoff of each Al layer, the sample is cleaned in a 250W downstream O$_{2}$ plasma asher for 10 min, both to remove resist residue and to enhance the native AlO$_{x}$ of the gates. After fabrication, the device is wire bonded to a (grounded) printed circuit board (PCB), and the on-chip shorting wires are cut with a diamond scribe attached to a micromanipulator. The PCB is then loaded into a dilution refrigerator at controlled humidity using ESD-safe tools wherever possible. Once at base temperature, the sample is illuminated with an infrared laser diode for $\sim$10 sec, which is observed to lower threshold voltages on the gates by several hundred mV.

\section{Detuning lever arms}
\label{sec:leverarms}

As described in the main text, the detuning lever arms are obtained from measurements of thermal broadening of the polarization lines. The detuning lever arms for both double dots are fit together using the measurement of the capacitive shift of the polarization lines. These capacitive shifts are equal for both double dots, in units of energy. In units of voltage, the magnitudes of the shifts differ due to the difference in lever arms for each double dot. Therefore, when fitting the thermal broadening data for detuning lever arms, we impose the additional constraint that the ratio of the lever arms should equal the observed ratio of the polarization line shifts in units of voltage. The uncertainty of the lever arms is calculated from the variance of the fit.

\begin{figure}
\includegraphics[width=0.48\textwidth]{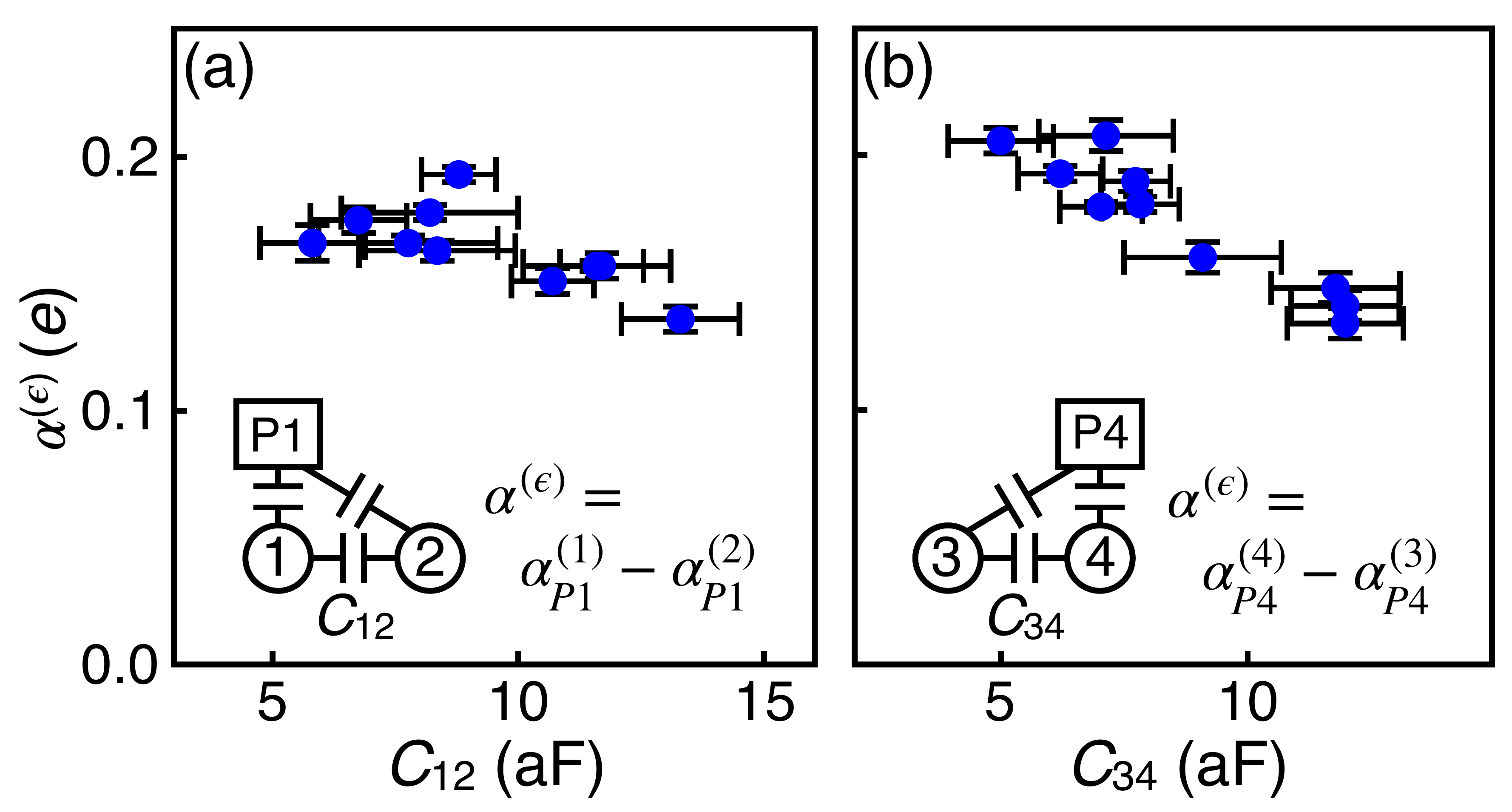}
\caption{
Detuning lever arms for (a) P1 and (b) P4 as a function of the intra-DD capacitance of the (a) left and (b) right double dot.
}
\label{fig:lever_arms}
\end{figure}

This procedure is repeated to measure the lever arms at every tuning of the barrier gate voltages. Since the detuning is the difference of the chemical potentials between two dots (e.g., $\epsilon_{12} = \mu_{1} - \mu_{2}$), the detuning lever arm for a particular gate is the difference between the lever arm for that gate and each dot, e.g., $\alpha_{P1}^{(\epsilon)} = \alpha_{P1}^{(1)} - \alpha_{P1}^{(2)}$. The detuning lever arm should therefore decrease when the two dots are pulled closer together and the action of a single gate on the two dots differs less. This effect is shown in Fig.~\ref{fig:lever_arms}, where detuning lever arms are plotted as a function of intra-DD capacitance for each double dot. The lever arms decrease by up to $36\%$ as intra-DD capacitance is increased, which confirms the importance of measuring the detuning lever arm after every change in intra-DD coupling so as to obtain accurate measurements of the double dot energetics at each tuning.

\section{Fitting for two-qubit Hamiltonian parameters}

\begin{figure}
\includegraphics[width=0.48\textwidth]{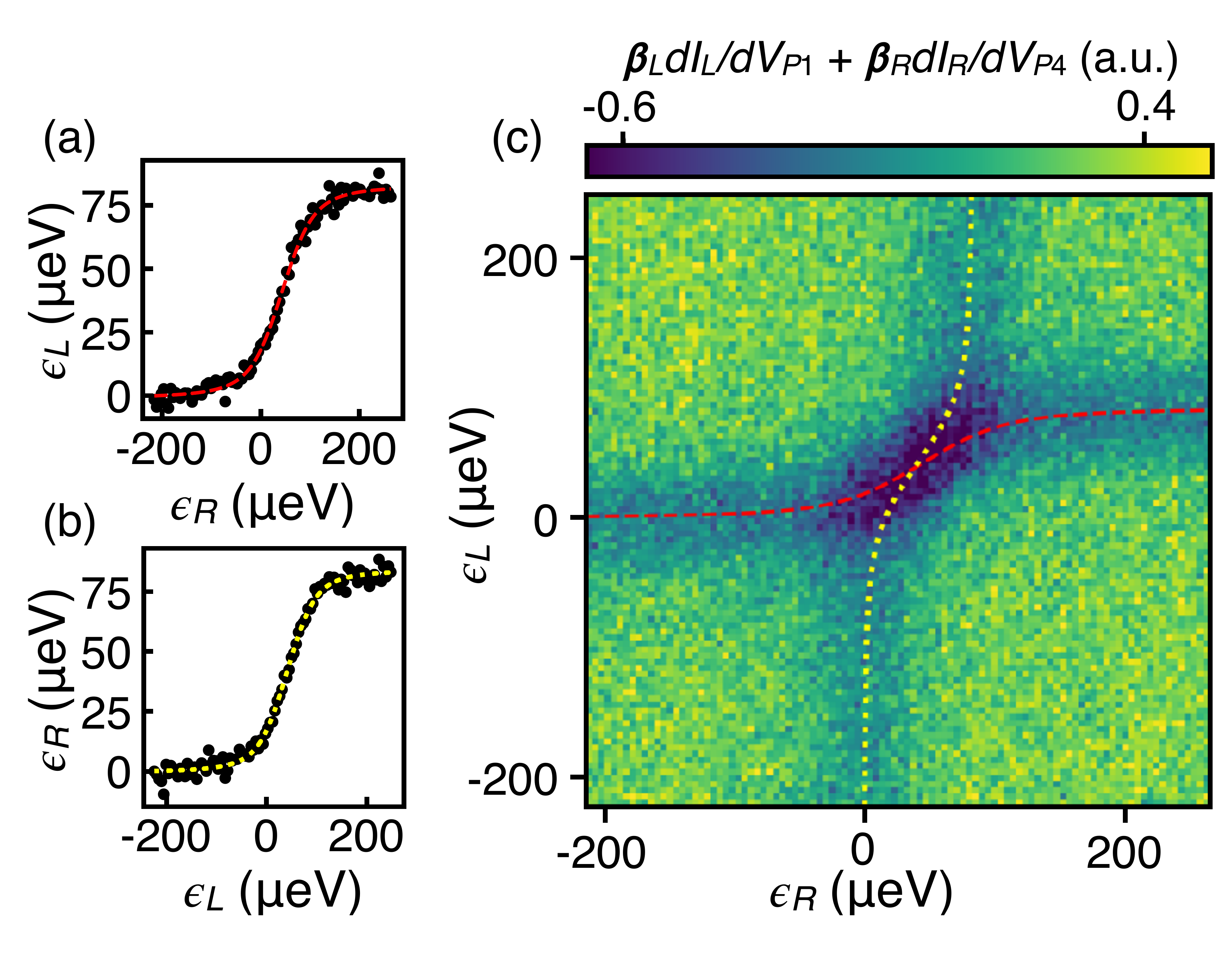}
\caption{
The location of the (a) left and (b) right double dot polarization line as it shifts due to the capacitive coupling. The dashed red and dotted yellow lines are the fits to the location data. (c) The full dataset from Fig. 3(a) in the main text with the fits to the polarization line locations overlaid.
}
\label{fig:g_t_fit}
\end{figure}

We fit the functions $P_{L(R)}$ (Eq. 2 of the main text) to the data using the following procedure. $P_{L(R)} \in [-1,1]$ are treated as functions of $\epsilon_{L}$ and $\epsilon_{R}$ and parametrized by $t_{L}$, $t_{R}$, and $g$ ($T_{e} = 155$ mK from the measurement described in section \ref{sec:leverarms}). $P_{L(R)} = 0$ corresponds to equal occupancy of both dots in the left (right) double dot, so we fit the roots of $P_{L(R)}$ to the locations of the left (right) polarization line in the data in Fig. 3(a) in the main text. The polarization line locations are obtained by fitting the derivative of a tanh function \cite{Dicarlo:2004p1440} to linecuts through the left (right) polarization lines and extracting the center points in $\epsilon_{L(R)}$. The resulting fit traces for each polarization line are shown in Fig.~\ref{fig:g_t_fit}(a) and (b) and overlaid on the 2D dataset in panel (c). The curves here are qualitatively similar to the tanh function that describes the charge polarization of a double dot as a function of its detuning \cite{Dicarlo:2004p1440} but in fact have a more complicated analytical form that depends on all the parameters of the coupled two-qubit system. Here only the curvature of the polarization lines is used in the fit, and the widths of the lines, which are also determined by $t_{L}$, $t_{R}$, and $T_{e}$, are not used. Including the widths as extra constraints would decrease the uncertainty of the fitted $t_{L}$ and $t_{R}$, but would add a significant computational overhead, so only the curvature fits in Fig.~\ref{fig:g_t_fit}(a) and (b) are used to extract the Hamiltonian parameters $t_{L}$, $t_{R}$ and $g$. 

The theoretical stability diagram shown in Fig. 3(b) of the main text incorporates the fitted polarization line curves with widths that are determined by the combined broadening from $t_{L}$, $t_{R}$, and $T_{e}$. The theoretical stability diagram has a background of zero, in contrast to the measured data, which has a background set by the transconductance of the charge sensors. We do not fit the background, since all the relevant information in the dataset is contained in the locations and dimensions of features in $\epsilon_{L}$ and $\epsilon_{R}$, and not in the absolute scale of the color (z) axis.

%

\section{Obtaining capacitances}

We measure the quantum dot charging energies $E_{Ci}$ and the electrostatic coupling energies $E_{Cij}$ by taking charge stability diagrams and measuring the offsets of the transition lines indicated in Fig. 1(d)-(f) in the main text. Specifically, we obtain $E_{Ci}$ from the distance in voltage space between the $N_{i}=2\rightarrow3$ and $N_{i}=3\rightarrow4$ transition lines and obtain $E_{Cij}$ from the distance in voltage space between the $N_{i}=2\rightarrow3$ transition line with $N_{j}=2$ and the $N_{i}=2\rightarrow3$ transition line with $N_{j}=3$, using appropriate lever arms. ($E_{Cij}=E_{Cji}$, so the two distances corresponding to this energy are averaged.) To fit lines to the transitions, we define a window around each transition of interest, use a peak-finding algorithm to generate a scatter plot within that window, and fit a line to the peak locations. We use the slopes of these lines to convert detuning lever arms to the lever arms between gates and their underlying dot chemical potentials. We use the offsets between the lines to obtain the electrostatic energies $E_{Ci}$ and $E_{Cij}$. The uncertainties of these energies are propagated from the uncertainties of the detuning lever arms and the variance of the linear fits to the transition lines.

\begin{figure}
\includegraphics[width=0.48\textwidth]{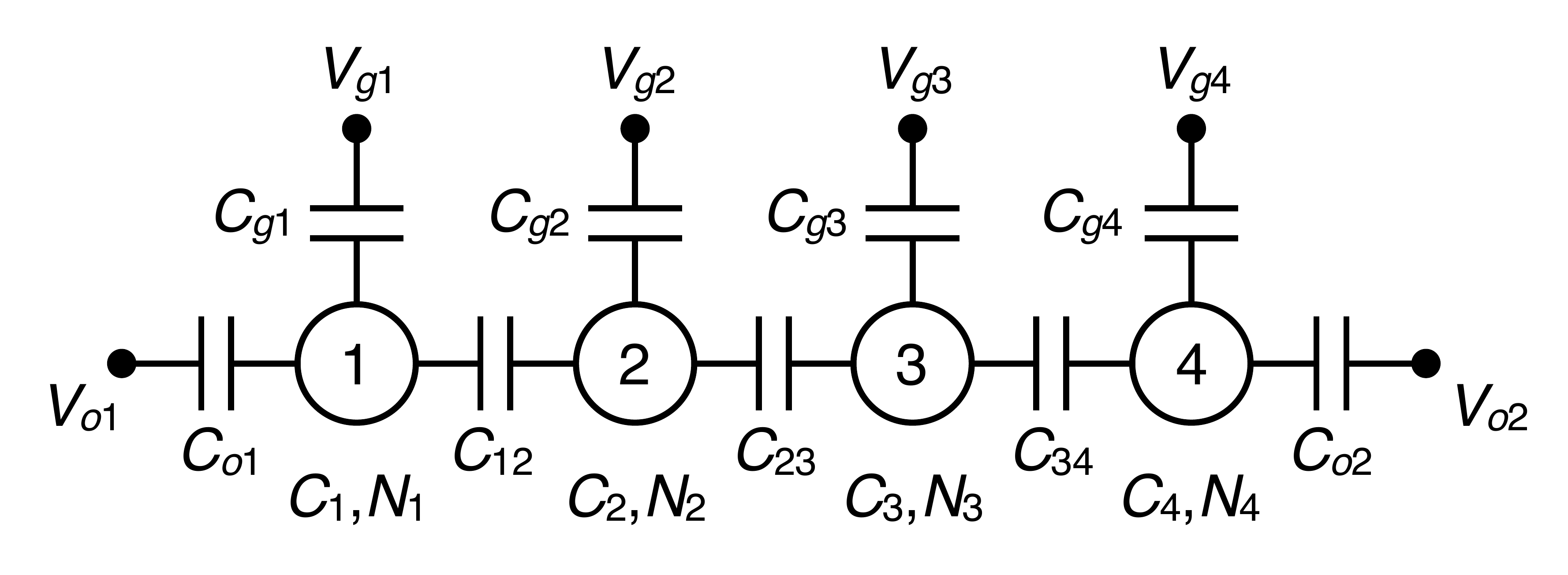}
\caption{
Diagram of the network of charge nodes joined by capacitors that we use to model the quadruple quantum dot array
}
\label{fig:C_network}
\end{figure}

We obtain the capacitances $C_{i}$ and $C_{ij}$ from the measured electrostatic energies using the following procedure. The full model of the network of charge nodes we use is shown in Fig.~\ref{fig:C_network}, with voltage sources from gates $V_{gi}$ and ohmic reservoirs $V_{oi}$ included. We calculate the electrostatic energy of the system \cite{vanderWiel:2003p1},
\begin{equation} 
U=\vec{Q} \cdot \mathbf{C}^{-1} \vec{Q},
\label{eq:U}
\end{equation}
using the definitions
\begin{equation}
\begin{split}
&\vec{Q} = 
\begin{pmatrix}
N_{1}e + C_{o1}V_{o1} + C_{g1}V_{g1} \\
N_{2}e + C_{g2}V_{g2} \\
N_{3}e + C_{g3}V_{g3} \\
N_{4}e + C_{o2}V_{o2} + C_{g4}V_{g4}
\end{pmatrix}
, \\
&\mathbf{C} =
\begin{pmatrix}
C_{1} & -C_{12} & 0 & 0 \\
-C_{12} & C_{2} & -C_{23} & 0 \\
0 & -C_{23} & C_{3} & -C_{34} \\
0 & 0 & -C_{34} & C_{4}
\end{pmatrix}
,
\end{split}
\label{eq:matrices}
\end{equation}
where $N_{i}$ is the number of electrons on node $i$. The resulting expression for $U$ contains terms proportional to $N_{i}^{2}$ and $N_{i}N_{j}$, whose coefficients equal $E_{Ci}$ and $E_{Cij}$, respectively. This yields the formulas for the energies:
\begin{equation}
E_{C1} = \frac{e^{2}}{|\mathbf{C}|} (C_{2}C_{3}C_{4} - C_{4}C_{23}^{2} - C_{2}C_{34}^{2}),
\end{equation}
\begin{equation}
E_{C2} = \frac{e^{2}}{|\mathbf{C}|} (C_{1}C_{3}C_{4} - C_{1}C_{34}^{2}),
\end{equation}
\begin{equation}
E_{C3} = \frac{e^{2}}{|\mathbf{C}|} (C_{1}C_{2}C_{4} - C_{4}C_{12}^{2}),
\end{equation}
\begin{equation}
E_{C4} = \frac{e^{2}}{|\mathbf{C}|} (C_{1}C_{2}C_{3} - C_{3}C_{12}^{2} - C_{1}C_{23}^{2}),
\end{equation}
\begin{equation}
E_{C12} = \frac{e^{2}}{|\mathbf{C}|} (C_{3}C_{4}C_{12} - C_{12}C_{34}^{2}),
\end{equation}
\begin{equation}
E_{C23} = \frac{e^{2}}{|\mathbf{C}|} (C_{1}C_{4}C_{23}),
\end{equation}
\begin{equation}
E_{C34} = \frac{e^{2}}{|\mathbf{C}|} (C_{1}C_{2}C_{34} - C_{34}C_{12}^{2}),
\end{equation}
where
\begin{multline}
|\mathbf{C}| = C_{1}C_{2}C_{3}C_{4} - C_{3}C_{4}C_{12}^{2} - C_{1}C_{2}C_{34}^{2} \\
- C_{1}C_{4}C_{23}^{2} + C_{12}^{2}C_{34}^{2}
.
\end{multline}
These formulas can be rearranged to give the capacitances in terms of the measured energies:
\begin{equation}
C_{1} = e^{2} \frac {E_{C2}} {E_{C1}E_{C2} - E_{C12}^{2}},
\end{equation}
\begin{equation}
C_{2} = e^{2} \frac {E_{C1}E_{C2}^{2}E_{C3} - E_{C12}^{2}E_{C23}^2}{E_{C2}(E_{C12}^{2} - E_{C1}E_{C2})(E_{C23}^{2} - E_{C2}E_{C3})},
\end{equation}
\begin{equation}
C_{3} = e^{2} \frac {E_{C2}E_{C3}^{2}E_{C4} - E_{C23}^{2}E_{C34}^2}{E_{C3}(E_{C23}^{2} - E_{C2}E_{C3})(E_{C34}^{2} - E_{C3}E_{C4})},
\end{equation}
\begin{equation}
C_{4} = e^{2} \frac {E_{C3}} {E_{C3}E_{C4} - E_{C34}^{2}},
\end{equation}
\begin{equation}
C_{12} = e^{2} \frac {E_{C12}} {E_{C1}E_{C2} - E_{C12}^{2}},
\end{equation}
\begin{equation}
C_{23} = e^{2} \frac {E_{C23}} {E_{C2}E_{C3} - E_{C23}^{2}},
\end{equation}
\begin{equation}
C_{34} = e^{2} \frac {E_{C34}} {E_{C3}E_{C4} - E_{C34}^{2}}
.
\end{equation}

The uncertainties of the capacitances are then propagated from the uncertainties of the electrostatic energies.


\section{Modeling the inter-dot capacitance}

\begin{figure}[t]
\includegraphics[width=0.48\textwidth]{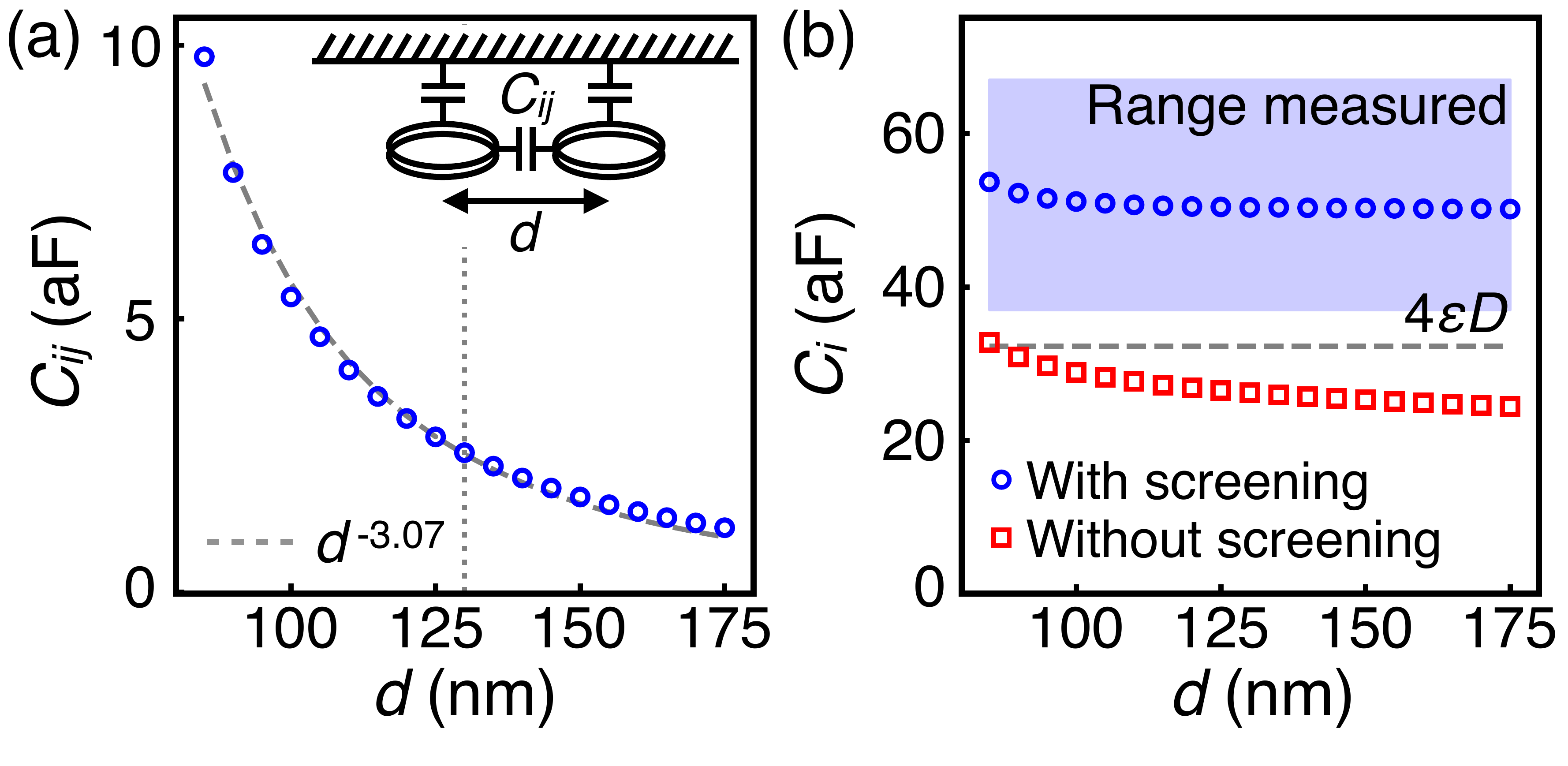}
\caption{
Calculations of the capacitance of quantum dots beneath a conducting plane. (a) Calculation of inter-dot capacitance, $C_{ij}$, as a function of the center-to-center distance, $d$, between the dots. The dashed line is a power law fit, with exponent -3.07. The vertical dotted line indicates the lithographic pitch of the plunger gates, 130 nm. (b) Calculations of total dot capacitance, $C_{i}$, as a function of $d$. Blue circles are calculated with screening effects from the metal gates. Red squares are calculated without screening. The horizontal dashed line corresponds to the expected self-capacitance of a disc of diameter $D=80$ nm, given by $4\epsilon D$. The light blue shaded region corresponds to the range of $C_{i}$ measured for the barrier gate voltages shown in Fig. 4 of the main text. 
}
\label{fig:C_sim}
\end{figure}

To estimate the variation in dot spacing associated with the variation in inter-dot capacitances observed in Fig. 4 in the main text, we model a double dot as a pair of conducting discs embedded in a semiconductor beneath a conducting plane, which incorporates the screening effects from the overlapping gate metal. The discs in the model have a diameter $D$ equal to the plunger gate width of 80 nm and a height of 1 nm. The distance of the dots below the conducting plane is 35 nm, which corresponds to the combined thickness of the SiGe spacer and the gate oxide in the device. COMSOL is used to model the system and to calculate the capacitance matrix. Fig.~\ref{fig:C_sim} shows the results of these calculations. In panel (a), the inter-dot capacitance, $C_{ij}$, is calculated as a function of the center-to-center distance, $d$, between the dots. The range over which $d$ is varied in these calculations is determined by the diameter of the dots in the model (80 nm) and the lithographic device pitch (130 nm) and is the range over which the distance between a pair of dots can vary inside the full array without overlapping each other or overlapping the neighboring dots. Over this range in $d$, $C_{ij}$ ranges from 1-10 aF, in good agreement with the range of inter-dot capacitances we observe in the device (2-13 aF). In panel (b), the self-capacitance of an individual dot, $C_{i}$, is calculated as a function of $d$. The blue circles are calculated with the screening effects from the gate metal included, and the red squares are calculated without screening effects. The $C_{i}$ values calculated without screening agree reasonably with the expected self-capacitance of a disc, given by $4\epsilon D$, marked on the plot by a horizontal dashed line. The $C_{i}$ values calculated with screening are systematically higher, and are in better agreement with the range of measured values for the points in Fig. 4 of the main text, indicated by the light blue shaded region. This agreement of both $C_{ij}$ and $C_{i}$ between the calculations and the measurements indicates that the dimension used here to model the dots ($D=80$ nm) is a reasonable approximation to the actual dot dimensions.

\section{Expressing capacitive coupling in terms of dot capacitances}
The capacitive coupling creates a detuning shift in one double dot due to a change in polarization of the other double dot:
\begin{multline}
g = \Delta \epsilon_{12} = \epsilon_{12}(N_{1},N_{2},N_{3}+1,N_{4}) \\
- \epsilon_{12}(N_{1},N_{2},N_{3},N_{4}+1),
\end{multline}
with an equivalent expression for $\Delta \epsilon_{34}$. The detuning is defined as the difference of quantum dot chemical potentials, which can be calculated from the electrostatic energy of the full capacitance network:
\begin{multline}
g = \Delta \epsilon_{12} = (U(0,1,1,0) - U(1,0,1,0)) \\
 - (U(0,1,0,1) - U(1,0,0,1))
 .
\end{multline}
Here we assume a total electron number of one per double dot, but the result is the same for higher fixed electron numbers. Expanding Eqs.~\ref{eq:U} and \ref{eq:matrices}, the terms proportional to source voltages and their capacitances cancel out, resulting in an expression that depends only on the seven capacitances that we find:
\begin{widetext}
\begin{equation}
g = \frac{e^{2}}{|\mathbf{C}|} C_{23}(C_{1}-C_{12})(C_{4}-C_{34})
= e^{2} \frac{C_{1}C_{4}C_{23}-C_{1}C_{23}C_{34}-C_{4}C_{12}C_{23}+C_{12}C_{23}C_{34}}{C_{1}C_{2}C_{3}C_{4}-C_{1}C_{2}C_{34}^{2}-C_{1}C_{4}C_{23}^{2}-C_{3}C_{4}C_{12}^{2}+C_{12}^{2}C_{34}^{2}}
.
\label{eq:coupling}
\end{equation}
\end{widetext}
This expression can be written as a series expansion in the reduced capacitances $c_{ij}=C_{ij}/C$ to obtain Eq. 4 in the main text. Eq. 4 also makes use of the approximation that all $C_{i}=C$. Across the tunings that we study here, the uniformity of the $C_{i}$'s varies, but they differ from the mean by at most $20\%$. The reduced capacitances are strictly $<1$. For $C \equiv \langle C_{i} \rangle$, $c_{12}$ and $c_{34}$ range from 0.1-0.2, and $c_{23}$ ranges from 0.03-0.08, putting any third or higher order terms in the expansion below $1\%$.

\bibliography{main}

\end{document}